# THE STOCHASTIC UNIVERSE:

# Professor A.M. Mathai's 75th Birthday

By Hans J. Haubold, United Nations, CMS Member since 1983

Abstract. A brief description and results of A.M. Mathai's research programme on statistics and probability, initiated in the 1970s, and its relation to physics is given.

**Introduction**

Professor A.M. Mathai, Director of the Centre for Mathematical Sciences India and Professor Emeritus of the Department of Mathematics and Statistics at McGill University Canada, initiated a research programme on statistics and probability for physics in the 1970s with the publication of three research monographs (see figures below) focusing on information theory and entropic functional (Jaynes, Shannon, Boltzmann-Gibbs), the normal distribution as emerging from random processes in statistics and physics (Gauss, Maxwell-Boltzmann), and the statistical characterization of random variables in terms of generalized hypergeometric functions (Meijer, Fox).

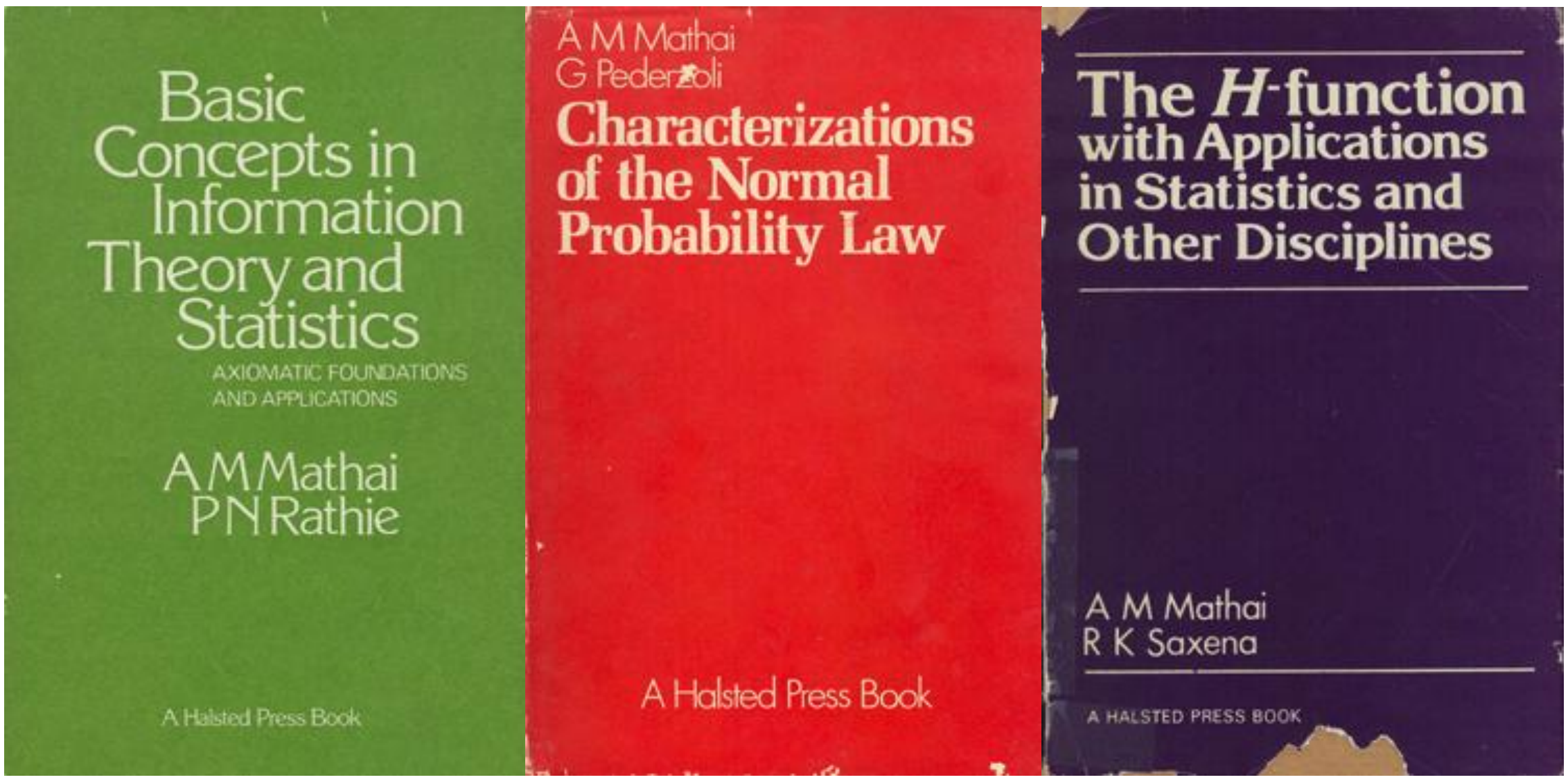

In past centuries, there have been remarkable relations between mathematics and physics (Manin 1981, Rompe and Treder 1979). Mathematics influenced the development of theoretical physics or problems arising from physics and vice versa theory and experiment in physics encouraged developments in mathematics and statistics. The history of special functions of mathematical physics is a testament for such developments.

**Mathematics and Statistics**

The classical subdivisions of mathematics are geometry, algebra, and analysis. Geometry studies a space made up of points with various sorts of structure. Analysis considers a function, possibly describing some physical quantity evolving in time, and its derivatives. Algebra handles a set of things with a law of composition. There is a fourth branch to the logic of mathematics still under hot debate how it is part of mathematics: Probability theory and statistical inference used frequently as foundations for scientific models (Uffink 2007). Carl Friedrich von Weizsaecker (2006), in his seminal study The Structure of Physics, elaborates that physics formulates general laws of nature in at least four forms emanating from the subdivisions of mathematics: (i) a family of functions, (ii) a differential equation, (iii) an extremum principle, and (iv) a symmetry group. Such general laws have been discovered and formulated in special and general relativity (Greene 2000) and quantum (field) theory (Veltman 2003), but not in statistical physics which developed not yet a set of generally accepted formal axioms, particularly for statistical thermodynamics and stochastic theory of non-equilibrium systems (Ebeling and Sokolov 2005). This is where Mathai's research programme is making contributions in one way or another: (i) standard and fractional differential equations for reaction, diffusion, reaction-diffusion, (ii) entropic, distributional, and differential pathways to beta-1, beta-2, and gamma families of distributions, and (iii) connection between fractional calculus and statistical distribution theory .

**Physics**

Concerning the interaction between physics and mathematics, twentieth century examples are Riemannian geometry in general relativity and the impact of quantum mechanics on the development of functional analysis. Einstein finalized general relativity in 1915, while quantum field theory has been an open frontier since its foundation in 1927 by Dirac based on Heisenberg's and Schroedinger's fundamental discoveries for quantum mechanics. This was 80-85 years ago. For the next fifty years there was not much interaction between theoretical physics and mathematics. Mathematics turned to more abstract accomplishments, while quantum field theory was formulated in a rather formal way. This changed in the mid-1970s when nonabelian gauge theories emerged as quantum field theories (Greene 2000).

The generally agreed perception is that there are two fundamental theories in twentieth century physics: general relativity and quantum field theory. General relativity describes gravitational forces on the scale of the macrocosmos (Greene 2000), while quantum field theory describes the interaction of elementary particles, electromagnetism, strong, and weak forces at the scale of the microcosmos (Veltman 2003). There remains to exist an inconsistency between the two theories. The formal quantization of general relativity leads to infinite formulas. Einstein invented general relativity to resolve an inconsistency between special relativity and Newtonian gravity (Pais 1982). Quantum field theory was invented to reconcile Maxwell's electromagnetism and special relativity with nonrelativistic quantum mechanics (Hoffmann 2010). But there were two basically different approaches. In Einstein's "thought experiments", which led to the discovery of general relativity, the logical framework came first. Then in Riemannian geometry of general relativity, the correct mathematical framework was found. In the development of quantum field theory on the other hand, there was no a priori conceptual basis; experimental clues played an important role, but there was no mathematical model. To date string theory is progressing in the pursuit to a formal quantization of general relativity: if accomplished it might be called a revolution in physics in the twenty first century.

A second revolution in the twenty first century may be the development of statistical mechanics beyond Maxwell, Boltzmann, and Gibbs, taking into account the need for a physical theory for stochastic theory of non-equilibrium systems (Prigogine 1980, Ebeling and Sokolov 2005). Indications are here that the mathematics for such a theory is given in fractional calculus with the H-function playing a central role in the solutions of stochastic fractional differential equations of the type of Liouville, master, Fokker-Planck, and reaction-diffusion (, Mandelbrot 1981, Klages et al. 2008, Honerkamp 1994, Tsallis 2009). In this regard, Mathai made contributions in his research programme. However, the physical interpretation of fractional time and space derivatives and integrals, applied to physical phenomena of reaction and diffusion, has not been discovered yet and a prospective physical theory based on a “Schroedinger equation for thermodynamics”, as referred to by Prigogine, even if he surely had in mind a type of master equation, fully incorporating the concept of entropic functionals, has not been discovered. Surely the H-function and fractional calculus seem to fit equations already known that govern spatio-temporal pattern formation which are central to a fully developed mathematical and physical theory for non-equilibrium systems.

A review of a large number of specific results that emanated in Mathai’s long teaching and research carrier in mathematics, statistics, and natural sciences is contained in two recently published books that have been prepared under his leading guidance:

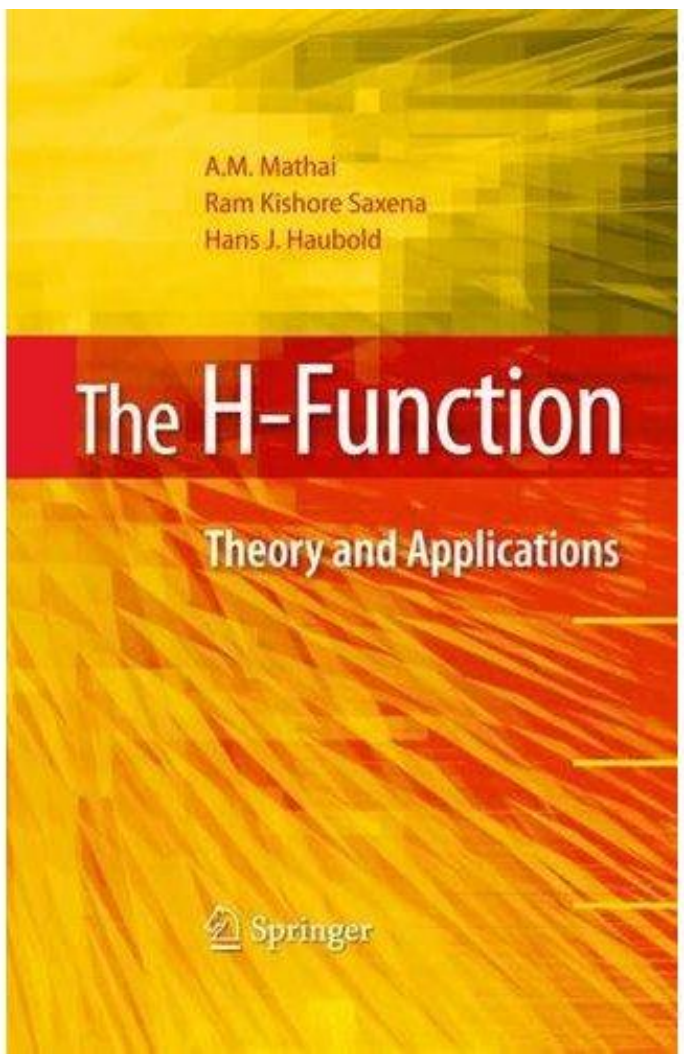

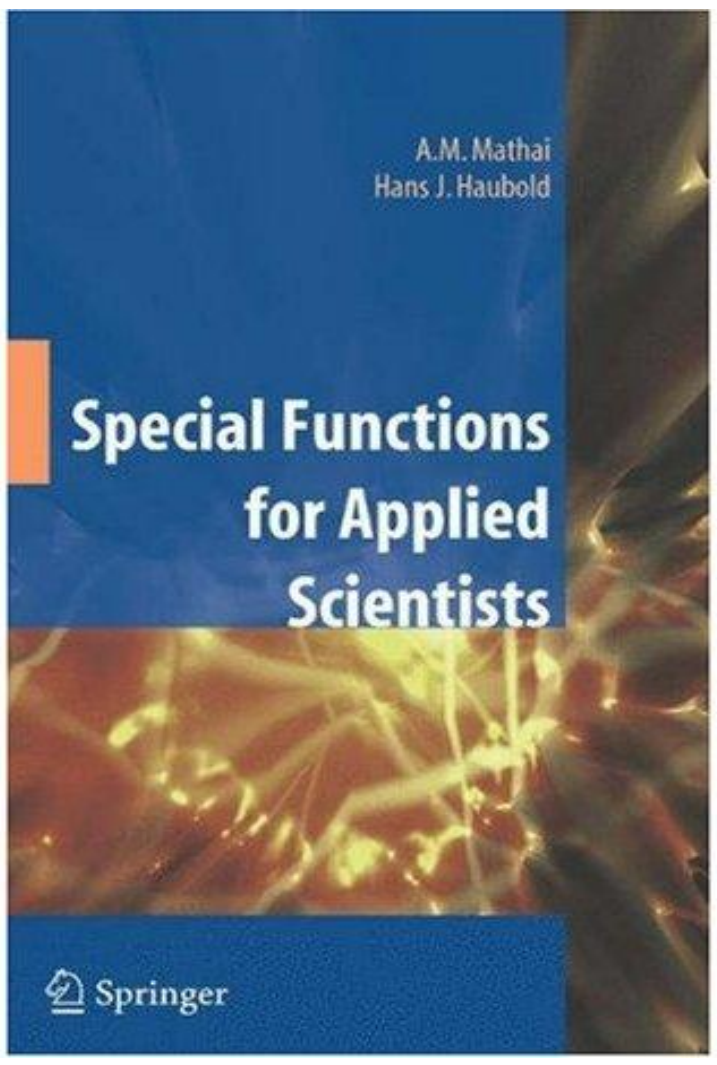

ICMS 2011: "International Conference on Mathematical Sciences in honour of Professor A.M.Mathai"

January 3-5, 2011

Venue: St Thomas College Pala, Kottayam – 686 574, Kerala, INDIA

www.stcp.ac.in

**International Organizing Committee**

Prof. H. J. Haubold (Chairman), *United Nations Office at Vienna (Austria)*
Prof. Thomas Mathew, University of *Maryland (USA)*
Prof. R. K. Saxena, *JNV University, Jodhpur (India)*
Prof. A. A. Kilbas, *Belarusian State University (Belarus)*
Prof. V. Kiryakov, *Bulgarian Academy of Sciences (Bulgaria)*
Prof. F. Mainardi, *University of Bologna (Italy)*
Prof. R. Gorenflo, *Free University of Berlin (Germany)*
Prof. H. M. Srivastava, *University of Victoria (Canada)*
Prof. S. B. Provost, *University of Western Ontario (Canada)*
Prof. P. Moschopoulos, *University of Texas at El Paso (USA)*
Prof. S. Kounias, *University of Athens (Greece)*
Prof. B. D. Acharya , *Formerly Advisor, DST, New Delhi (India)*
Prof. R. Natarajan, University of *Lakehead (Canada)*

**National Organizing Committee**

Prof. K. K. Jose (Chairman), *Mahatma Gandhi University, Kottayam*
Prof. Peeyush Chandra, *IIT Kanpur, Kanpur*
Prof. Ashish Sengupta, *ISI Kolkata, Kolkata*
Prof. B. K. Dass, *University of Delhi, New Delhi*
Prof. A. K. Singh, *SERC Division, DST, New Delhi*
Prof. A. K. Agarwal, *Punjab University, Chandigarh*
Prof. Umesh Singh, *Banaras Hindu University, Varanasi*
Prof. Debasis Kundu, *IIT Kanpur, Kanpur*
Prof. R. K. Kumbhat, *JNV University, Jodhpur*
Prof. A. Sukumaran Nair, *Mahatma Gandhi University, Kottayam*

**Local Organizing Committee**

Joy Jacob (Chairman)
K. M. Kurian
Alex Thannippara
Sebastian George
Benny Kurian
Seemon Thomas (Co-ordinator)
Dilip Kumar
Vishnudas V.

**Enquiries can be addressed to:**

Seemon Thomas, Mob: 91-9495325341
seemonpala@rediffmail.com

Joy Jacob, Mob:91-9446126315
jjstc2000@yahoo.com

Dilip Kumar, Mob: 91-9446195433
dilipkumar.cms@gmail.com

St Thomas College Pala is one of the first institutions in India to introduce masters program in Statistics. To celebrate the 75th birth anniversary of Professor A.M. Mathai who is one of our prestigious alumni and former faculty, we organize an international conference in his honour during January 3-5, 2011. For a cv of Professor A.M. Mathai log on to www.math.mcgill.ca/mathai/.
The International Conference on Mathematical Sciences (ICMS 2011) aims to bring together academic scientists and researchers to exchange and share their experiences and research findings in Mathematical Sciences, and discuss the practical challenges encountered and the solutions adopted. To promote international participation of researchers from outside India, foreign experts are proposed as invited speakers.
The section titles of ICMS-2011 include but are not limited to:
Integral transforms and special functions; Differential equations and applications; Integral, difference, functional equations and fractional calculus; Real and complex analysis; Applied problems of analysis; Theoretical and applied problems of mechanics; Astrophysics; Distribution theory; Stochastic processes; Statistical inference; Multivariate analysis; Mathematical and stochastic modeling; Computation and simulation.

**Call for Abstracts**

The organizers will accept papers for presentation at the conference subject to approval by referees. Please send abstracts electronically (preferably in LaTeX or Word format) to Dr. Joy Jacob at the email address jjstc2000@yahoo.com. The title of the abstract must be followed by the name(s) of the author(s) (please underline the name of the presenter), their affiliation(s) and e-mail address (es), the body of the abstract, AMS classification numbers, and up to five keywords.
Deadline for submission of abstract: 31 August 2010.

**Paper Submission**

All full papers will be peer reviewed and chosen based on originality, content, correctness, relevance to conference, contributions and readability. Prospective authors are kindly invited to submit full text including results, tables, figures and references. Full text (.doc, .tex with .pdf) will be accepted only by electronic submission through seemonpala@rediffmail.com or jjstc2000@yahoo.com.
Deadline for submission of full manuscript: 30 September 2010.

**Special Journal Issue**

All submitted papers in ICMS 2011 will have opportunities for consideration for a Special Issue of a reputed international journal. The selection will be carried out during the review process as well as at the conference presentation stage. Submitted papers must not be under consideration by any other journal for publication. The final decision will be made based on peer review reports by the guest editors and the Editor-in-Chief jointly.

**Important Dates**

| | |
|---|---|
| Submission of abstract | by August 31, 2010 |
| Submission of full manuscript | by September 30, 2010 |
| Notification of acceptance for presentation | by October 15, 2010 |
| Conference dates | January 3-5, 2011 |

**Registration**

All participants will have to be registered and the registration form in word format is available from www.stcp.ac.in. Early registration is recommended since the number of participants will be limited.

| Registration Fees | For foreign participants | For Indian participants |
|---|---|---|
| By October 31, 2010 | US $150.00 | Rs 1200 |
| Between November 1, 2010 and December 31, 2010 | US $175.00 | Rs 1300 |
| After December 31, 2010 | US $200.00 | Rs 1500 |

Students and local participants who do not require accommodation are allowed a reduction of US $50 or Rs 400 in the above tariffs. Registration fee includes:
Food and moderate accommodation in the guest house during conference days, conference materials, banquet and entertainment programs.
Registration fee can be transferred electronically to: A/c- Co-ordinator ICMS 2011, Account No. 0453053000006856, South Indian Bank, Arunapuram, India. For transfer in India (RTGS or NEFT) use IFSC Code: SIBL0000453. For international transfer use SWIFT Code: SOININ 55. Electronic transfer of fee must be indicated with full reference in the registration form.
Please send completed registration form through email or by post to: Seemon Thomas, Co-ordinator, ICMS 2011, Department of Statistics, St. Thomas College, Pala-686 574, INDIA, Ph: +91-4822-201288, Fax: +91-4822-216313, Email: seemonpala@rediffmail.com, seemon@stcp.ac.in.

Presentation certificate will be issued to those who present papers, and all others will be issued attendance certificate provided they attend all sessions.

Technical Equipments for Electronic PowerPoint/Acrobat Presentations will be available. For more details regarding the conference log on to our website www.stcp.ac.in.
Information concerning hotels and their charges will be provided later. We also plan to have a cultural event, and a short sightseeing trip.
For details regarding nearby places of tourist interest log on to www.keralatourism.org.